\documentclass[twocolumn,prl,floats,superscriptaddress]{revtex4-2}
\usepackage[LGR,T1]{fontenc}
\usepackage{textcomp}
\usepackage[utf8]{inputenc}
\usepackage{verbatim}
\usepackage{amsmath}
\usepackage{amssymb}
\usepackage{graphicx}
\usepackage[version=4]{mhchem}
\usepackage[bookmarks=false,
 breaklinks=false,pdfborder={0 0 1}, colorlinks=true]
 {hyperref}

 \makeatletter

\usepackage{bm}
\usepackage{array}
\usepackage{xcolor}
\usepackage{float}
\usepackage{color}
\usepackage{xspace}
\usepackage{multirow}
\usepackage{dcolumn}

\graphicspath{{./figures/}{./}}

\newcommand{\AgI}{\ce{AgI}\xspace}
\newcommand{\AgPO}{\ce{AgPO_3}\xspace}
\newcommand{\AgIAgPO}{\AgI -\AgPO}
\newcommand{\AgIxAgPOmx}{\ce{(AgI)_{x}(AgPO_3)_{1-x}}\xspace}

\makeatother

\begin{document}

\title{Revealing the origin of ionic conduction in silver-iodide-doped silver phosphate glass}

\author{J. Freedberg}
\email{fredbrg2@illinois.edu}
\affiliation{Department of Physics, The Grainger College of Engineering, University of Illinois Urbana-Champaign, Urbana, Illinois 61801, USA}
\affiliation{Materials Research Laboratory, The Grainger College of Engineering, University of Illinois Urbana-Champaign, Urbana, Illinois 61801, USA}

\author{J. Maduzia}
\affiliation{Department of Mechanical Science and Engineering, The Grainger College of Engineering, University of Illinois Urbana-Champaign, Urbana, Illinois 61801, USA}

\author{A. Santoso}
\affiliation{Department of Physics, The Grainger College of Engineering, University of Illinois Urbana-Champaign, Urbana, Illinois 61801, USA}
\affiliation{Materials Research Laboratory, The Grainger College of Engineering, University of Illinois Urbana-Champaign, Urbana, Illinois 61801, USA}

\author{R. Singh}
\affiliation{Department of Mechanical Science and Engineering, The Grainger College of Engineering, University of Illinois Urbana-Champaign, Urbana, Illinois 61801, USA}

\author{P. Ferreira}
\affiliation{Department of Mechanical Science and Engineering, The Grainger College of Engineering, University of Illinois Urbana-Champaign, Urbana, Illinois 61801, USA}

\author{F. Mahmood}
\email{fahad@illinois.edu}
\affiliation{Department of Physics, The Grainger College of Engineering, University of Illinois Urbana-Champaign, Urbana, Illinois 61801, USA}
\affiliation{Materials Research Laboratory, The Grainger College of Engineering, University of Illinois Urbana-Champaign, Urbana, Illinois 61801, USA}

\date{September 6, 2026}

\begin{abstract}
Fast ionic transport is a defining feature of many solid electrolytes, yet its microscopic origin is not fully understood. In the absence of microscopic insights, the development of next-generation solid-state batteries remains largely empirical. Most existing measurements access either the low-frequency transport response or the high-frequency bound polarization, yet the intermediate mesoscopic frequency regime is where ionic transport emerges. By varying the \AgI concentration ($x$) and performing time-domain terahertz spectroscopy (TDTS) in a prototypical glassy electrolyte \AgIxAgPOmx, we reveal this intermediate frequency regime and identify a crossover from bound-current-dominated conduction to conductivity arising from short-range dispersive ionic transport. We find that bound polarization associated with the bond-bending motion of the \ce{P\text{-}O^-\text{-}Ag^+} motif is present across compositions but is insufficient to produce ionic transport on its own. Transport emerges only when this polarization is embedded in a sufficiently soft \AgPO glassy matrix and accompanied by a high carrier density. These ingredients together take the system from a vibrationally bound response to short-range dispersive motion.
\end{abstract}
\maketitle

In solid electrolytes, charge is carried by mobile ions embedded in a host network rather than via free electrons \cite{berzinaRoleSolidElectrolyte2010, sunFocusedReviewStructures2023}. Thus, carrying an ionic current requires a rearrangement of the material itself. This motion spans a broad hierarchy of timescales, from DC transport which is indicative of long-range diffusion to high-frequency dynamics associated with the vibrational modes of the network \cite{jonscherUniversaldielectricResponse1977, singhSuperlinearFrequencyDependence2016, dyreUniversalityAcConduction2000a, sidebottomUniversalApproachScaling1999a}. While there must be a crossover between these two dynamical limits, how exactly the localized, vibration-dominated response becomes long-range diffusive transport remains unresolved  \cite{dyreFundamentalQuestionsRelating2009, m.b.m.mangionFastIonicConduction1987, sidebottomUniversalApproachScaling1999a}.

As candidates for solid-state batteries, solid electrolytes promise improved safety, mechanical stability, and energy density compared to liquid lithium-ion batteries \cite{zhuIntermediateTemperatureFuel1994}. Realizing that promise, however, depends on achieving high ionic conductivity in a solid, and thus requires resolving the emergence of diffusive transport.

\begin{figure}
    \centering
    \includegraphics[width = 1\linewidth]{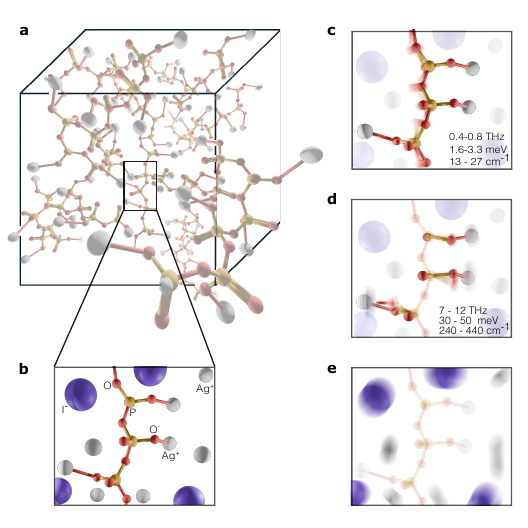}
    \caption{\textbf{Coupled host and ionic dynamics in \AgIAgPO}. (a) Schematic of the \AgPO glass network, with the highlighted region enlarged in (b) to show the local structure and mobile $\mathrm{Ag^+}$ and \ce{I^-} ions. (c--e) Schematic representations of the three current contributions relevant to this work: (c) bound current associated with low-frequency motion of the \AgPO backbone, (d) bound current associated with the local pendulum-like motion of the \ce{P\text{-}O^-\text{-}Ag^+} motif, and (e) free current associated with ionic transport. Motion in (c--e) is illustrated by blurred copies of the equilibrium structures. \label{fig:cartoon}}
\end{figure}

Solid electrolytes can have radically different chemistry, structure, and proposed transport mechanisms, ranging from vacancy hopping to motion between localized sites to coordination changes. Descriptions of ionic conduction often share a common tendency to treat the host as static and the ions as the main actors \cite{morimotoMicroscopicIonMigration2019,zhuIntermediateTemperatureFuel1994}. However, the static-host picture cannot fully explain the necessary ingredients for strong ionic conductivity. Within the past year, simulations spanning diverse material classes suggest that vibrational modes facilitate ionic transport \cite{wooFrontiersDefectTransport2026, xuFastIonicTransport2026b, liLocalStructureDictates2026, liFastIsotropicLiIon2026b, XiaolinElasticPropertiesAmorphous2026}. Recent experiments have likewise associated the activation of soft modes with enhanced conductivity in the well-known crystalline ionic conductors \ce{Li_7La_3Zr_2O_{12}} and \ce{Li_{0.5}La_{0.5}TiO_3} \cite{linRoleTHzPhonons2025,phamCorrelatedTerahertzPhonon2026}. Meanwhile, other THz-scale studies in single- and polycrystalline fluorite oxides reported similar crossover-like behavior, but interpreted that response primarily in terms of ion dynamics rather than an active role of the host lattice \cite{morimotoMicroscopicIonMigration2019}. This convergence across distinct material classes points to a broader principle in which the host structure is an active participant in the emergence of ionic transport, rather than a static backdrop.

The prototypical glassy fast-ion conductor \AgIxAgPOmx provides a particularly useful platform for testing this hypothesis \cite{nakayamaGlassStructureEnergetic1994,jacobsDirectEbeamWriting2016, hirokitakahashiStructuralStudySuperionic1994, bondarevFluctuationTheoryLow1999b, novitaElasticFlexibilityFastion2009}. Here, mobile charge carriers are supplied by the \AgI dopant, while the structural framework is provided by the amorphous \AgPO host matrix. This glassy network exhibits low-frequency vibrational modes--the so-called boson peak \cite{schirmacherHeterogeneousElasticityTale2022}--centered near 0.6~THz \cite{novitaElasticFlexibilityFastion2009, fontanaLowtemperatureOpticalThermal1999, pallesVibrationalSpectroscopicBond2016}. Although these modes have long been hypothesized to influence ionic transport \cite{awanoFarInfraredMillimeterWave2010, morimotoMicroscopicIonMigration2019,poletayevPersistenceMemoryIonic2024a,hanayaActivationPropertiesAg+ion2003}, their role in the current response has not been directly measured. Combined with the absence of grain boundaries, the presence of a single mobile ionic species, and the ability to systematically vary dopant concentration, \AgIxAgPOmx provides a model system for isolating the essential ingredients of ionic motion in a solid.

Experimentally, ionic motion has largely been accessed through two classes of measurements \cite{lestanguennecFrequencydependentIonicConductivity1994a, tsonosExploringHighFrequencies2022}. In the low-frequency limit, measurements emphasize ionic transport in the free conductivity, $\mathbf{J}_{\mathrm{free}}$, as seen in Refs.~\cite{riveraCationMassDependence2002,ngaiCageDecayConstant2002,dyreUniversalityAcConduction2000a, lunkenheimerResponseDisorderedMatter2003}. At high frequencies, the vibrational response of the lattice or host matrix gives rise to a time-dependent polarization $\mathbf{P}$, whose dynamics constitute  a bound current. Measurements relying on the bound current are shown for example in Refs.~\cite{novitaElasticFlexibilityFastion2009,velliStructuralInvestigationMetaphosphate2005,fontanaLowtemperatureOpticalThermal1999}.  Electrodynamically, both terms participate in the total current $ \mathbf{J}_{\mathrm{tot}}$, such that $ \mathbf{J}_{\mathrm{tot}}= \mathbf{J}_{\mathrm{free}} \, +\,\frac{\partial \mathbf{P}}{\partial t}$ \cite{sokolovOpticalPropertiesMetals1967}.  However, because these experimental approaches operate in widely separated frequency regimes, it has historically been difficult to establish how structural dynamics manifest in the current response, or how they give rise to transport-like behavior. At low frequencies the bound current is small, while at sufficiently high frequencies the free current becomes small. Between these limits, neither contribution can be ignored and the bound current may therefore play a defining role by shaping the electrodynamic environment governing the emergence of ionic transport. Resolving this intermediate regime requires an experimental approach that directly probes the total current response on the intrinsic timescale of ionic motion, where polarization and transport contributions coexist. This mesoscopic regime lies in the terahertz (THz) frequency range, making it a natural window in which to examine how vibrational dynamics and transport-like response become intertwined in disordered ionic systems.

In this work, we use time-domain terahertz spectroscopy (TDTS) to probe \AgIAgPO across the relevant microscopic timescales for the first time. We directly observe a crossover between dispersive hopping and vibrational dynamics, but only at high \AgI concentrations, where the phosphate backbone becomes sufficiently flexible \cite{micoulautFastionConductionFlexibility2009,novitaElasticFlexibilityFastion2009, funkeCorrelatedIonicHopping2005} and the carrier density is high enough to support fast motion. Our results show that the \AgPO network is an active contributor to ionic motion rather than a passive spectator. These distinct pathways are summarized schematically in Fig. \ref{fig:cartoon}. Ionic transport (Fig.~\ref{fig:cartoon}(e)) arises when vibrations of the mechanically softened backbone (Fig.~\ref{fig:cartoon}(c)) drive a coupled local bound response (Fig.~\ref{fig:cartoon}(d)) that sets mobile $\mathrm{Ag^+}$ ions into motion.

Our measurements access the intrinsic timescales of ionic motion, providing a direct view of the microscopic processes governing ionic conduction. We find that fast ionic transport emerges only when mobile ions occupy a mechanically soft, glassy network containing a polarizable local motif. Only when all three conditions are met does the local bound response evolve into short-range dispersive ionic motion. These results in \AgIAgPO suggest that strong ionic conduction at nonzero frequency may therefore be understood as a property of the coupled ion-host system rather than of mobile carriers moving through a passive background, underscoring that electrolyte design must account for the electrodynamic and mechanical environment of ionic motion.\\

\begin{figure}
    \centering
    \includegraphics[width = 1\linewidth]{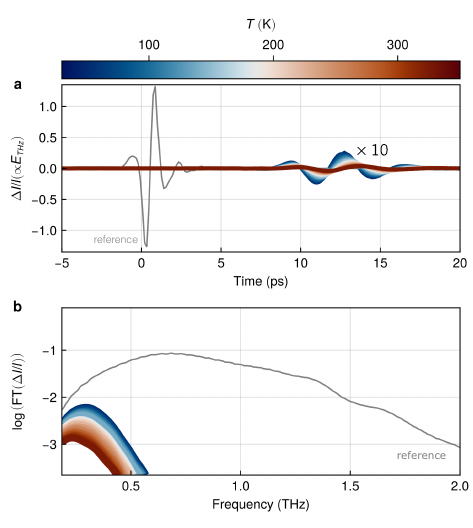}
    \caption{\textbf{Time- and frequency-domain THz response of the 20\% AgI composition at various temperatures.} (a) Transmitted THz electric field as a function of time at different temperatures (blue to red), together with the reference measurement through an open aperture (gray). (b) Corresponding frequency-domain spectra obtained by Fourier transforming the time-domain traces in (a). Transmission through the sample suppresses the THz bandwidth relative to the reference. \label{fig:raw_data}}
\end{figure}

\begin{figure*}
    \centering
    \includegraphics[width = 1\linewidth]{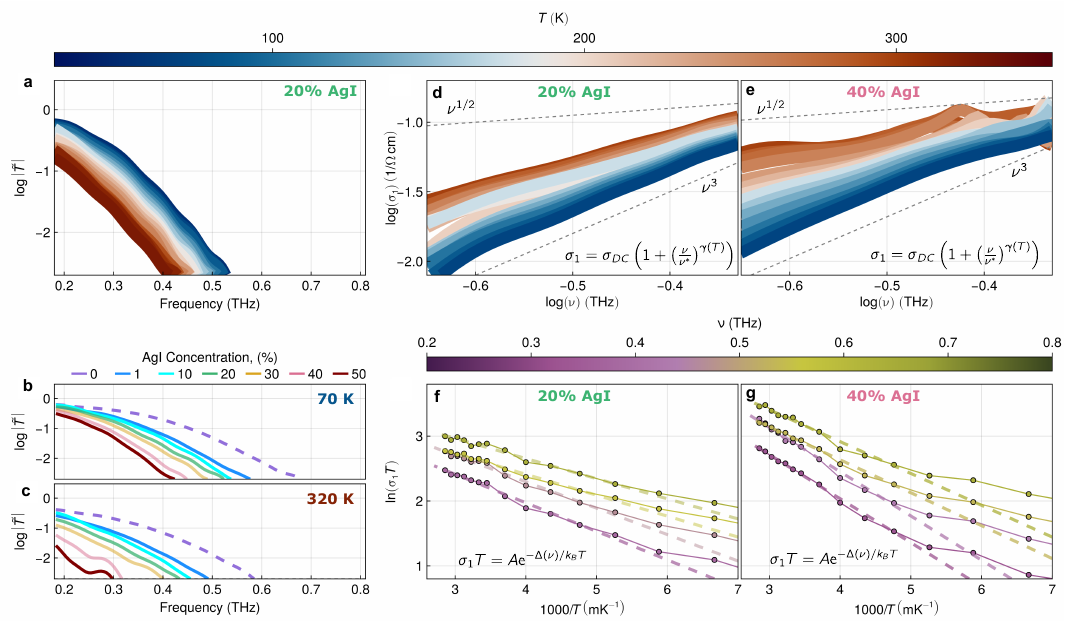}
    \caption{\textbf{Temperature-, frequency-, and concentration-dependent THz transmission and conductivity of \AgIAgPO}. (a) Magnitude of the complex THz transmission for a representative \AgI concentration as a function of frequency and temperature. (b,c) THz transmission as a function of frequency and \AgI concentration at fixed temperatures. (d,e) Real part of the complex conductivity, $\sigma_1(\nu,T)$, as a function of frequency for representative \AgI concentrations at different temperatures. Dashed lines indicate powers of one-half and three. (f,g) Arrhenius representation of the conductivity for representative concentrations at selected frequencies, used to extract the activation energy $\Delta$ from the high-temperature slopes.\label{fig:analysis}}
\end{figure*}

\noindent\textbf{THz transmission and absorption edge}\\
Fig.~\ref{fig:raw_data}(a) shows the time-domain THz traces for the reference measurement without the sample (gray) and for transmission through the sample at different temperatures (blue to red). Fourier transforming these traces yields the frequency-domain spectra shown in Fig.~\ref{fig:raw_data}(b), from which the THz transmission is obtained. Fig.~\ref{fig:analysis}(a) shows the transmission magnitude as a function of temperature for a fixed \AgI concentration of 20\%, while Fig.~\ref{fig:analysis}(b,c) shows the corresponding concentration dependence at fixed temperatures.

\begin{figure*}
    \centering
    \includegraphics[width = 1\linewidth]{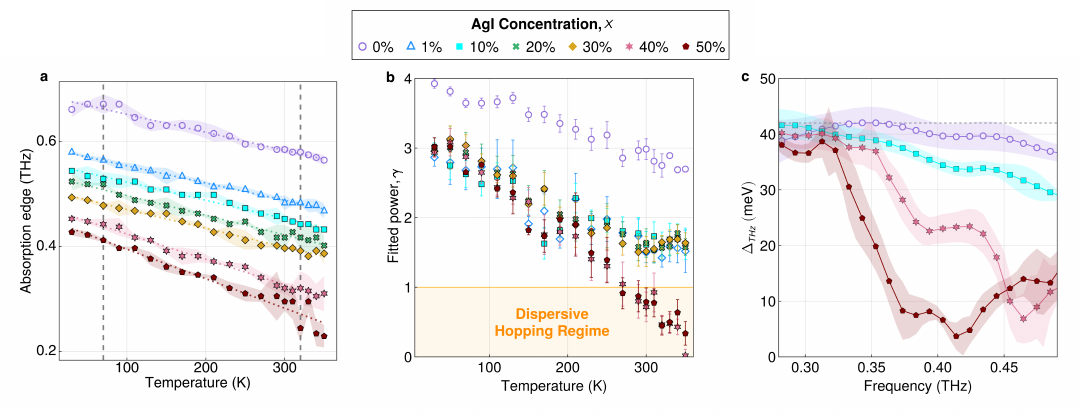}
    \caption{\textbf{THz response separates rigid and flexible elastic regimes in \AgIAgPO.} Non-percolating compositions are shown with open markers, while percolating compositions, including both rigid and flexible glasses, are shown with filled markers. (a) Onset frequency of the THz absorption edge, defined as the frequency at which the transmission falls below the experimental noise floor. Vertical lines indicate the two temperatures shown in Fig.~\ref{fig:analysis}(b,c). (b) Exponent $\gamma$ extracted from fits to Eq.~\ref{eq:power_law}. (c) Frequency-dependent Arrhenius barriers as a function of frequency for four representative compositions; the complete concentration dependence is shown in the Supplemental Information. \label{fig:results}}
\end{figure*}

As shown in Fig.~\ref{fig:analysis}(a-c), all compositions, \textit{including} the undoped \AgPO glass, exhibit a pronounced absorption edge within the THz window, well within our measurement bandwidth (Fig.~\ref{fig:raw_data}(b)). The onset frequency is determined from the intersection of the transmission with the experimental noise floor and is plotted as a function of temperature in Fig. \ref{fig:results}(a). Previous work has attributed this feature to low-frequency vibrational modes of the \AgPO backbone \cite{novitaElasticFlexibilityFastion2009, novitaMolecularStructureAgPO32005}. Upon \AgI doping, the onset frequency shifts to lower values, but remains near $\sim0.6$~THz across all compositions.\\

\noindent\textbf{THz conductivity in frequency}\\
To characterize the frequency-dependent charge response, we extract the real part of the complex conductivity, $\sigma_1(\nu,T)$, from the measured THz transmission by numerically inverting the Fresnel equations \cite{sokolovOpticalPropertiesMetals1967}. This quantity contains contributions from both free and bound currents. The resulting conductivity is shown in Fig.~\ref{fig:analysis}(d,e) as a function of temperature for different \AgI concentrations. To isolate the \AgI-induced response, we subtract the bound-current contribution of undoped \AgPO from the higher-concentration samples. Details of the inversion and subtraction procedures are provided in the Supplemental Information.

We first analyze the frequency dependence of $\sigma_1(\nu,T)$ at fixed temperature. For solid electrolytes such as \AgIAgPO, this frequency response is conventionally described through three dynamical regimes \cite{tsonosExploringHighFrequencies2022}. Across these regimes, the conductivity follows a piecewise power law consistent with the universal scaling described by Sidebottom and others \cite{jonscherUniversaldielectricResponse1977, singhSuperlinearFrequencyDependence2016, morimotoMicroscopicIonMigration2019, sidebottomInfluenceCationConstriction2000, sidebottomUniversalApproachScaling1999a}
\begin{equation}
    \sigma_1(\nu; \,T ) = \sigma_{DC}\left(1+ \left(\frac{\nu}{\nu^*}\right)^\gamma\right), \label{eq:power_law}
\end{equation}

\noindent where $\sigma_{DC}$ is the DC conductivity, $\nu^* = \nu^*(T)$ is the ``attempt'' frequency of the conducting ion, and $\gamma = \gamma(T)$ is the scaling exponent. At low frequencies, ions can fully follow the applied field, producing a DC conductivity plateau ($\gamma = 0$). In \AgIAgPO, this regime occurs below tens of kHz \cite{singhSuperlinearFrequencyDependence2016, micoulautFastionConductionFlexibility2009}. At higher frequencies, the response enters an intermediate regime dominated by short-range hopping. Here, $0<\gamma<1$, characteristic of the dispersive regime first described by Jonscher \cite{jonscherUniversaldielectricResponse1977, nakayamaGlassStructureEnergetic1994}. At still higher frequencies, even short hops are suppressed, and ions undergo localized vibrations within the glassy potential landscape \cite{morimotoMicroscopicIonMigration2019, tsonosExploringHighFrequencies2022, singhSuperlinearFrequencyDependence2016}. In \AgIAgPO, this is expected to occur near 0.6~THz \cite{novitaElasticFlexibilityFastion2009, novitaMolecularStructureAgPO32005}. Thus, extracting $\gamma$ allows us to identify the dynamical regime probed by the THz response.

We next determine which compositions enter the dispersive hopping regime within the THz bandwidth. To do so, we fit the power law described by Eq.~\ref{eq:power_law} to the conductivity data shown in Fig.~\ref{fig:analysis}(d,e). In Fig.~\ref{fig:results}(b), we plot the fitted exponent $\gamma$ as a function of temperature and concentration. Below $200~\mathrm{K}$, all compositions exhibit $\gamma>1$, consistent with a superlinear, vibration-dominated response.  Above $200~\mathrm{K}$, the data separate cleanly according to elastic regime \cite{novitaElasticFlexibilityFastion2009, novitaMolecularStructureAgPO32005, micoulautFastionConductionFlexibility2009}: rigid and intermediate (``percolating'') compositions remain at $\gamma>1$ across the full temperature range, while only the flexible, high-$\mathrm{AgI}$ glasses continue to evolve into the dispersive hopping regime. Specifically, only compositions in the flexible elastic regime ($x\geq37.5\%$) cross from $\gamma>1$ to $\gamma<1$ within the THz window. \\

\noindent\textbf{THz conductivity with temperature}\\
Having established how the frequency dependence evolves across the different regimes, we next examine the temperature dependence at fixed frequency to identify the characteristic energy scales governing the THz response. We fit the high-temperature conductivity to an Arrhenius form \cite{morimotoMicroscopicIonMigration2019, sidebottomInfluenceCationConstriction2000,micoulautFastionConductionFlexibility2009,singhSuperlinearFrequencyDependence2016}
\begin{equation}
    \sigma_1(T; \,\nu)T = A\mathrm{e}^{-\Delta /k_B T},\label{eq:arrhenius}
\end{equation}

\noindent where $A = A(\nu)$ is a fitted prefactor, $\Delta = \Delta(\nu)$ is the barrier height, and $k_B$ is Boltzmann's constant. 

Fig.~\ref{fig:analysis}(f,g) show the Arrhenius behavior for two representative concentrations, from which we extract the barrier height $\Delta$ using Eq. \ref{eq:arrhenius}. The resulting activation energies are plotted as a function of frequency for four representative compositions in Fig.~\ref{fig:results}(c), with additional concentrations shown in the Supplemental Information. Across the THz range, $\Delta$ is approximately 40~meV for all compositions, nearly an order of magnitude smaller than the corresponding DC transport barriers \cite{rodriguesChargeCarrierMobility2011, liuMechanicalVsElectrical1986, micoulautFastionConductionFlexibility2009}. This reduced energy scale suggests that the THz response probes local, low-energy motion within the glassy potential landscape rather than long-range ionic transport. Approaching the absorption edge, $\Delta$ decreases further, with substantially stronger softening observed in the flexible compositions.\\

\noindent\textbf{Implications for the origin of ionic transport}\\
Our data demonstrate that neither a high concentration of $\mathrm{AgI}$ nor a flexible $\mathrm{AgPO_3}$ backbone alone can produce fast ionic transport. Instead, strong ionic conductivity emerges from the combination of a high density of mobile $\mathrm{Ag^+}$ ions and soft vibrational modes of the glassy network. This conclusion is supported by three observations from the same set of temperature-dependent measurements: (i) the selective entry of high-$\mathrm{AgI}$ glasses into the dispersive hopping regime within the THz window, (ii) the temperature- and concentration-dependent onset of the low-frequency backbone vibrational modes, and (iii) THz activation energies that follow the softening of the backbone rather than the DC transport barrier.

Building on earlier work that identified rigid, intermediate (“percolating”), and flexible elastic regimes in \AgIxAgPOmx glasses using thermodynamic, vibrational, and DC transport measurements \cite{novitaElasticFlexibilityFastion2009, novitaMolecularStructureAgPO32005, micoulautFastionConductionFlexibility2009}, we adopt this elastic classification as the organizing framework for our analysis. Here, we show that this classification does not only correlate with DC transport trends, but also determines whether vibrational dynamics can evolve into transport-like ionic motion on the THz timescale. Because THz measurements are sensitive to \textit{both} bound current arising from local polarization and free current associated with transport on the intrinsic ionic timescale, they provide a direct means to disentangle the mechanisms responsible for ionic transport.

Across the different analyses in Fig.~\ref{fig:results}, the data separate cleanly into rigid and flexible regimes. This separation coincides with the transition to a flexible glassy network rather than with the formation of a percolating \AgI cluster. Although the \AgI percolation threshold occurs near 9\%, all compositions below 40\% exhibit quantitatively similar behavior and remain distinct from the 40\% and 50\% samples. Thus, while mobile ions are necessary for ionic transport, the sharp change observed in our data is governed by the elastic state of the glassy host network.

The role of the backbone is evident in the temperature and concentration dependence of the absorption edge. Fig.~\ref{fig:results}(a) shows that this response evolves more strongly with temperature in the flexible compositions, while the presence of the absorption edge even in undoped \AgPO demonstrates that backbone motion alone is insufficient to produce ionic transport. Consistent with this distinction, Fig.~\ref{fig:results}(b) shows that only compositions with a flexible host matrix enter the dispersive hopping regime on the THz timescale.

The interplay between these vibrational and transport responses is further revealed by the frequency-dependent activation barriers in Fig.~\ref{fig:results}(c). As the absorption edge is approached, the apparent activation barrier decreases for all compositions, but the reduction is substantially stronger in the flexible, high-\AgI glasses, where it approaches zero near the edge. Below the absorption edge, the characteristic energy scale is approximately $40~\mathrm{meV}$, closely matching the energy of the \ce{P\text{-}O^-\text{-}Ag^+} bond-bending mode identified by Raman spectroscopy \cite{novitaElasticFlexibilityFastion2009}. This local motion, illustrated schematically in Fig.~\ref{fig:cartoon}(d), produces a bound polarization current whose energy scale is indicated by the dotted horizontal line in Fig.~\ref{fig:results}(c). The correspondence between this bound response and the reduction of the transport barrier supports a picture in which local polarization and backbone softening together facilitate ionic motion.

Taken together, these measurements show that fast ionic transport in \AgIAgPO emerges from the coupled dynamics of mobile $\mathrm{Ag^+}$ ions and the glassy host network. High carrier density alone is insufficient: transport-like motion appears only when the backbone is sufficiently flexible and the local polarization response is active. By resolving the temperature, frequency, and composition dependence of this crossover, THz spectroscopy directly accesses the intermediate dynamical regime connecting vibrationally bound response to dispersive ionic motion.

\section*{Methods}

Time-domain terahertz spectroscopy (TDTS) measurements were performed using a custom-built transmission setup driven by a Yb:KGW amplifier laser (PHAROS, LIGHT CONVERSION) operating at a center wavelength of 1030~nm and a repetition rate of 50~kHz, with a pulse duration of approximately 160~fs. THz radiation was generated by optical rectification in N-benzyl-2-methyl-4-nitroaniline (BNA) and incident normally on the sample. The transmitted THz electric field was detected using electro-optic sampling in a \ce{CdTe} (110) crystal   \cite{nahataCoherentDetectionFreely1996}. A synchronized near-infrared gate pulse was spatially overlapped with the transmitted THz pulse at the electro-optic crystal, and their relative delay was controlled using a motorized delay stage. Measurements were performed in the linear-response regime using weak peak THz fields.

Seven different $\mathrm{AgI}$ compositions were studied: 0\%, 1\%, 10\%, 20\%, 30\%, 40\%, and 50\%. The material was heated to above the glass transition temperature, which is \AgI concentration dependent, and pressed to the desired thickness. This synthesis route has been extensively characterized across the \AgIAgPO composition range in prior work \cite{novitaElasticFlexibilityFastion2009, jacobsPaintingDirectWriting2015}, including structural, thermal, and transport measurements. The transport properties measured here were consistent with these established composition-dependent trends. The material was stored in the dark in either a desiccator or a vacuum vessel since \AgIAgPO is hygroscopic and light sensitive. To protect the integrity of the samples during testing, they were stored in a nitrogen purge box in a covered container. The samples were mounted in a Janis SHI-950 cryostat, which can reach temperatures from 4~-~400~K. Further information about time-domain THz measurements and analysis can be found in the Supplemental Information.

Data analysis was carried out in the programming language Julia \cite{Julia-2017} using the plotting library Makie \cite{Makie_2021}. The \AgPO structure in Fig. \ref{fig:cartoon} was generated with Avogadro2 \cite{hanwell2012avogadro} and Blender \cite{blender} using the molecular structure from Ref. \cite{AgPO3_cif}.

\vspace{2mm}
\noindent\textbf{Acknowledgments:}
We thank E. Ertekin and N. Perry for useful discussions. J.F. thanks B. Shklovskii, E. D. Dahlberg, P. Crowell, R. Aquino, and J. Vi\~{n}als for insightful questions. J.F. also gratefully acknowledges W. J. Meese for many valuable discussions throughout this project. 
This work was supported by the Illinois Materials Research Science and Engineering Center, funded by the National Science Foundation MRSEC program under NSF award number DMR-2309037.
F.M. acknowledges support from the EPiQS program of the Gordon and Betty Moore Foundation, Grant GBMF11069 and NSF CAREER Award number DMR-2144256.
J.F.~and F.M.~acknowledge support from the Center for Emergent Materials, an NSF MRSEC, under Grant No.~DMR-2011876.
J.F. acknowledges support from the Joe Greene Postdoctoral Fellowship from the Materials Research Laboratory at the University of Illinois, Urbana-Champaign.

\vspace{2mm}
\noindent\textbf{Author contributions:} J.F. and A.S. performed the THz experiments and J.F. completed the corresponding analysis. J.M. and R.S. synthesized and characterized the samples. J.F. and F.M. wrote the manuscript with input from all of the authors. F.M. and P.F. conceived and supervised this project.

\vspace{2mm}
\noindent\textbf{Competing interests:} The authors declare no competing interests.

\vspace{2mm}
\noindent\textbf{Data availability:} The data in this manuscript are available from the corresponding authors upon reasonable request.

\vspace{2mm}
\noindent\textbf{Corresponding authors:} Correspondence to Jennifer Freedberg ~(fredbrg2@illinois.edu) and/or Fahad Mahmood~(fahad@illinois.edu).

\bibliographystyle{unsrt}
\bibliography{newnames_paper_refs.bib}

\end{document}